\documentclass[%
        12pt,
        notitlepage, 
        tightenlines,
        eqsecnum,
        floats,
        floatfix,
        nofootinbib,
        amsmath,amssymb,
        aps,prd,
        superscriptaddress,
        longbibliography]{revtex4-2}

\usepackage[utf8]{inputenc}
\DeclareUnicodeCharacter{0302}{}
\usepackage{microtype}
\usepackage{verbatim}
\usepackage{mathtools}
\usepackage{graphicx}
\usepackage[caption=false]{subfig}
\usepackage{float}
\usepackage{amssymb}
\usepackage{xcolor}

\usepackage[hidelinks]{hyperref}
\usepackage{cleveref}
\def\dd{{\rm d}}

\definecolor{newgreen}{rgb}{0.0, 0.75, 0.0}

\begin{document}

\title{Effective Kerr geometry from loop quantum gravity}

\author{Francesco Fazzini}
\email{francesco.fazzini@unb.ca}
\affiliation{Department of Mathematics and Statistics, University of New Brunswick, Fredericton, NB, Canada E3B 5A3}

\begin{abstract}

We derive an effective Kerr metric from an effective Schwarzschild metric inspired by loop quantum gravity through the Newman-Janis algorithm. The resulting spacetime is free from the classical ring singularity and does not allow the existence of closed time-like curves, due to a lower radial bound in its domain inherited by the spherically symmetric seed metric. We give a possible interpretation to this lower bound by casting the metric in generalized Painlevé-Gullstrand coordinates.
We analyze the horizon structure and the ergoregion pointing out the similarities and the differences with the classical Kerr spacetime. The study of the trajectory of the zero angular momentum observer allows to show two novel effects due purely to quantum gravity related to the frame dragging and the repulsive behaviour in the deep quantum region. We derive an effective separation Carter constant and the geodesic equations. Finally, we specialize the geodesic analysis to equatorial circular trajectories obtaining  a modified third Kepler law and the equation defining the photon sphere.

\end{abstract}

\maketitle

\section{Introduction}

Black holes are a striking prediction of Einstein theory of gravity, which existence has been proved by an increasing number of direct and indirect observations. Although they are nowadays largely accepted by the scientific community, some of their main features like physical singularities and existence of time-like closed curves (for the axisymmetric case) are considered failures of Einstein theory.\par
The common belief is that a quantum theory of gravity should be able to cure such pathological features, once quantum effects are kept in account. 
Among the proposed theories of quantum gravity, loop quantum gravity (LQG) seems mature enough to address these kind of issues, giving at the effective level regular geometries (with the correct classical limit) exhibiting quantum corrections in the strong curvature regime that allow to avoid singularities. \par
The spherically symmetric case (effective Schwarzschild solution) has been extensively studied in the framework of loop quantum gravity in the past decade. Many quantitatively different effective metrics have been derived from effective equations of motion, guessed by reasonable physical considerations or obtained from junction conditions \cite{DeLorenzo:2014pta,Modesto:2004xx,Ashtekar:2005qt,Boehmer:2007ket,Campiglia:2007pb,Chiou:2008nm,Brannlund:2008iw,Corichi:2015xia,Cortez:2017alh,Olmedo:2017lvt,BenAchour:2018khr,Ashtekar:2018cay,Alesci:2019pbs,Assanioussi:2019twp,Gambini:2008dy,Gambini:2013ooa,Husain:2004yz,Chiou:2012pg,Gambini:2020nsf,Hossenfelder:2009fc,Rovelli:2014cta,deCesare:2020swb,Belfaqih:2024vfk,Alonso-Bardaji:2023vtl,Alonso-Bardaji:2023fcv,Alonso-Bardaji:2021yls,Cafaro:2024vrw,Han:2023wxg}.
For reviews, see \cite{Perez:2017cmj,Gambini:2022hxr,Ashtekar:2023cod}.
Even if such models can differ from each other in significant predictions like the number of horizons, they all agree in the avoidance of the classical Schwarzschild singularity. \par 
Despite the huge amount of literature regarding the spherical symmetric case, very little is done for the axisymmetric one. To the knowledge of the Author no effective equations are available in the axisymmetric case so far.\par
In the classical theory, the only tool available to derive the Kerr solution without solving Einstein equations is the Newman-Janis algorithm \cite{Newman:1965tw}. Requiring only a spherically symmetric seed metric (not even a solution of any field equations) as input, it is the easiest way to generalize the Kerr (and possibly the Kerr-Newman) solution to their effective counterparts.\par
There are results in this direction already available in literature, which use  LQG-based spherically symmetric seed metrics to derive and analyze effective Kerr metrics \cite{Brahma:2020eos,Muniz:2024wiv,Kumar:2022vfg,Suresh:2024hth,Islam:2022wck,Dong:2024alq,Chen:2024sbc}. This is the path that we follow in the present paper: we derive a modified Kerr metric from a particular effective Schwarzschild metric widely studied and used in literature (see e.g. \cite{Kelly:2020uwj,Lewandowski:2022zce,Fazzini:2023scu,Giesel:2023hys,Han:2023wxg,Munch:2021oqn,Munch:2021oqn,Parvizi:2021ekr,Kelly:2020lec,Han:2023wxg,Giesel:2021rky}). The metric we are going to present has been already introduced in \cite{Brahma:2020eos}, even if not analyzed in detail. \par
It is worth emphasizing the limitations of the approach adopted: in the classical theory one can check that the output of the algorithm is actually a solution of Einstein field equations. 
Here, since we do not have effective equations of motion at our disposal we do not have an a priori warranty that the output metric is the solution of any effective equations. However, due to the extremely simple applicability of the algorithm it seems worth to explore this direction, with the idea of capturing at least qualitatively the main features of an equations of motion (EOM)-based effective metric.\par
The main results of this paper are: the analysis of the avoidance of the classical ring singularity through the Generalized Painlevé-Gullstrand (GPG) coordinates and absence of closed time-like curves; the derivation of the horizon and ergoregion structures; the discovery of two novel effects, namely the \emph{frame braking} and \emph{radial braking}, that arise in the high curvature regime and are absent in the classical solution;
the derivation of an effective Carter constant and geodesic equations; the generalization of the third Kepler law and the determination of the governing equation for the photon sphere. \par
We use units where $c=1$ in this manuscript.

\section{The seed metric and its domain lower bound}
\label{s.seed}
The spherical symmetric seed metric that we use as input for the Newman-Janis algorithm has been derived from LQG-based effective equations of motion in \cite{Kelly:2020uwj,Giesel:2023hys}. In advanced Painlevé-Gullstrand coordinates it takes the following form:
\begin{equation}
 ds^2=-\dd t^2+(\dd r+N^{r} \dd t)^2+r^2 \dd \Omega^2,
 \label{PG}
\end{equation}
where $N^r$ is the radial component (and the only non-vanishing one) of the shift vector, which reads:
\begin{equation}
 N^r=\sqrt{\frac{R_{S}}{r}\left( 1-\frac{\gamma^2 \Delta R_{S}}{r^3}\right)}.
\end{equation}
Here $R_S=2GM$ is the usual Schwarzschild radius, $\Delta \propto l_{P}^2$ is the minimum area gap in LQG and $\gamma$ is the Barbero-Immirzi parameter.\par
A relevant feature of this line element is that, differently from its classical counterpart, the domain of the shift vector has a lower bound at $r_{min}\equiv (\gamma^2 \Delta R_S)^{\frac{1}{3}}$. Consequently the line element does not contain the classical singularity at $r=0$ but it does not cover entirely the spatial slices $\Sigma(t)$. This could signal a failure of Painlevé-Gullstrand (PG) coordinates (see \cite{Faraoni:2020ehi} for an analysis of a similar issue in the Reissner-Nördstrom solution), since in a different gauge like the Schwarzschild one this bound disappears:
\begin{equation}
 ds^2=-\left(1-\frac{R_{S}}{r}+\frac{\gamma^2 \Delta R_{S}^2}{r^4} \right)\dd t^2 +\frac{1}{1-\frac{R_{S}}{r}+\frac{\gamma^2 \Delta R_{S}^2}{r^4}}\dd r^2 +r^2 \dd \Omega^2 ~. 
 \label{sch}
\end{equation}
This metric contains a time-like physical singularity like the Reissner-Nordström spacetime and no lower bound is present. However, this interpretation turns out to be wrong if one includes matter in the picture:
any initial collapsing dust profile in the marginally bound case with total gravitational mass $M$ cannot be compressed in an areal radius smaller than $r_{min}$ by gravity \cite{Fazzini:2023ova,Kelly:2020lec, Giesel:2023hys}. In the non-marginally bound case such lower bound is slightly modified due to breakdown of Birkhoff theorem at the effective level \cite{Cipriani:2024nhx, Cafaro:2024vrw}. \par
The peculiarity of the PG gauge is that it makes the physical lower boundary of the vacuum region manifest: as well known, the PG time is the proper time of a free-falling time-like geodesic observer, and his velocity, due to the repulsive character of quantum gravity, becomes zero when the observer reaches the radius $r_{min}$ \cite{Kelly:2020uwj}. The same happens however to the areal radius of the outermost physical shell of any matter distribution of compact support with the same total gravitational mass, making PG coordinates ideal to show this lower bound of the effective vacuum metric.

\section{Application of the Janis-Newman algorithm}
\label{s.cosm}
In order to apply the Newman-Janis algorithm to the metric \eqref{PG} we need to cast it in retarded Eddington-Finkelstein coordinates \cite{Drake:1998gf}. It can be easily checked that this yelds to the following line element:
\begin{equation}
\dd s^2=-\left(1-\frac{R_S}{r}+\frac{\gamma^2\Delta R_S^2}{r^4}     \right)\dd u^2-2\dd u\dd r+r^2\dd \Omega^2 ,
\label{1}
\end{equation}
    where, for the considerations made in the previous section $r \in [r_{min},+\infty)$.\par 
    The first step of the algorithm consists in a tetrad decomposition of the metric in terms of four null vectors $l^\mu, n^\mu, m^\mu, \Bar{m}^\mu$ (here $m$ and $\Bar{m} $ are complex-valued and conjugate to each other):
\begin{equation}
g^{\mu \nu}=-l^\mu n^\nu  - n^\mu l^\nu +m^\mu \Bar{m}^\nu +\Bar{m}^\mu m^\nu ,
\label{2}
\end{equation}
related by:
\begin{align}
& l^{\mu}l_{\mu}=n^\mu n_\mu=m_{\mu}m^{\mu}=\Bar{m}^{\mu}\Bar{m}_{\mu}=0  ~,\label{3}    \\
& l_{\mu}n^{\mu}=-m_{\mu}\Bar{m}^{\mu}=-1~ , \label{4}    \\
&l_{\mu}m^{\mu}=n_{\mu}m^{\mu}=l_{\mu}\bar{m}^{\mu}=n_{\mu}\Bar{m}^{\mu}=0~. \label{5}
\end{align}
To proceed, we need to find the explicit expression of these null vectors for our line element using \eqref{3}--\eqref{5}. Recalling that $u=const.$ are hypersurfaces on which the trajectories of outgoing null rays lie, by calling $\lambda$ the affine parameter for outgoing null geodesics we have ${\dd u}/{ \dd \lambda}=0$. If we call $l^\mu$ the four-velocity of outgoing radial null geodesics then ~$l^{\mu}=(0,f,0,0)$; we will fix $f$ in a moment. For ingoing null rays, the normalization of the four-velocity ($n^\mu n_\mu$=0) gives
\begin{equation}
    n^{\mu}=\left(1,-\frac{1}{2} \left( 1-\frac{R_S}{r}+\frac{\gamma^2\Delta R_S^2}{r^4} \right),0,0\right).
\end{equation}
The condition \eqref{4} fixes $f$ to 1, and this concludes the construction of the first two null vectors. The other two four-vectors will have only $\theta$ and $\phi$ components to ensure the orthogonality conditions \eqref{5}, and they can be built using the so called non-holonomic method:
\begin{align}
 & m_{\mu}\dd x^{\mu}=\frac{1}{\sqrt{2}}(\sqrt{g_{\theta \theta}}\dd \theta+i\sqrt{g_{\phi \phi}}\dd \phi )~, \\
 &\Bar{m}_{\mu}\dd x^{\mu}=\frac{1}{\sqrt{2}}(\sqrt{g_{\theta \theta}}\dd \theta-i\sqrt{g_{\phi \phi}}\dd \phi )~.
\end{align}
Therefore
\begin{align}
& m^{\mu}=\left(0,0,\frac{1}{\sqrt{2}r},\frac{i}{\sqrt{2}r \sin\theta} \right), \\
&\bar{m}^{\mu}=\left(0,0,\frac{1}{\sqrt{2}r},-\frac{i}{\sqrt{2}r \sin\theta} \right).
\end{align}
It can be easily check that they satisfy \eqref{2}--\eqref{5}. The four-vectors introduced above form a null tetrad, that we call: $Z^{\mu}_{a}\equiv (l^{\mu},n^{\mu},m^{\mu},\Bar{m}^{\mu})$. \par
The next step is to apply a complex transformation to the null tetrad:
\begin{align}
 &x^\rho \rightarrow \Tilde{x}^\rho=x^\rho+i y^\rho(x^\sigma)~,
 \label{firsttransf}\\
 &Z^{\mu}_{a}\rightarrow \Tilde{Z}^{\mu}_{a}(\Tilde{x}^{\rho},\Tilde{\Bar{x}}^{\rho})~,
 \label{newz}
\end{align}
where $y$ is a real function of $x$. The transformation is constrained by two requirements: the new metric constructed out of the new tetrad has real components, and both the tetrad field and the metrics become the original ones if $\Tilde{x}^{a}$ are real:
\begin{equation}
\Tilde{Z}^{\mu}_{a}(\Tilde{x},\Tilde{\Bar{x}})|_{\Tilde{x}=\tilde{\bar{x}}}=Z^{\mu}_{a}(x)~. 
\label{16}
\end{equation}
 Notice that the transformation \eqref{newz} generally is not a coordinate transformation for the tetrad and more importantly is not unique, since once \eqref{firsttransf} is fixed there are many transformations of the form \eqref{newz} that satisfy the two constraints stated above. The choice adopted in our case is the following:
\begin{align}
 &l^{\mu}\rightarrow \Tilde{l}^{\mu}=\delta^{\mu}_{1}~,  \label{17} \\
&n^{\mu} \rightarrow \Tilde{n}^{\mu}=\delta^{\mu}_{0}-\frac{1}{2}\left[1-\frac{R_S}{2}\left(\frac{1}{\Tilde{r}}+\frac{1}{\Tilde{\Bar{r}}} \right)+\frac{\gamma^{2}\Delta R_{S}^{2}}{\Tilde{r}\Tilde{\Bar{r}}Re(\Tilde{r})^{2}} \right]\delta^{\mu}_{1} ~,\label{18} \\
& m^{\mu}\rightarrow \Tilde{m}^{\mu}=\frac{1}{\sqrt{2}\Tilde{r}}\left( \delta^{\mu}_{2}+\frac{i}{\sin{\Tilde{\theta}}}\delta^{\mu}_{3}\right)~,\label{19}
\end{align}
while the transformation \eqref{firsttransf}:
\begin{equation}
\begin{cases}
  &\Tilde{t}=t      \\
  &\Tilde{r}=r+i a \cos\theta \\
  &\Tilde{\theta}=\theta \\
  &\Tilde{\phi}=\phi
\end{cases}
\end{equation}
where $a$ at this stage is an undetermined constant. Writing the new tetrad in terms of the old variables, we get
\begin{align}
&\Tilde{l}^{\mu}=\delta^{\mu}_{1}~, \\
& \Tilde{n}^{\mu}=\left(1,-\frac{1}{2}\left(1-\frac{rR_{S}}{r^{2}+a^{2}\cos^{2}\theta}+\frac{\gamma^{2}\Delta R_{S}^{2}}{r^{2}(r^{2}+a^{2}\cos^{2}\theta)}   \right),0,0    \right) ~,   \\
& \Tilde{m}^{\mu}=\frac{1}{\sqrt{2}(r-ia\cos\theta)}\left(ia \sin\theta,-ia\sin\theta,1,\frac{i}{\sin\theta} \right) ~,  \\
& \Tilde{\Bar{m}}^{\mu}=\frac{1}{\sqrt{2}(r+ia\cos\theta)}\left(-ia \sin\theta,ia\sin\theta,1,-\frac{i}{\sin\theta} \right)~.
\end{align}

The condition \eqref{16} can be immediately checked for this transformation, while the condition on the real-valued metric components can be checked by reconstructing the new metric from $\Tilde{Z}^{\mu}_{a}$. A simple computation gives
\begin{equation}
g_{\mu \nu}=
\begin{pmatrix}
-1+D      & -1 & 0 & -{a D\sin^{2}\theta} \\
    -1     & 0 & 0 & a\sin^{2}\theta \\
   0 & 0 & \rho^{2} & 0 \\
-{a D \sin^{2}\theta}& a\sin^{2}\theta & 0& \Sigma \sin^2\theta
\label{kerrcoord}
\end{pmatrix}
\end{equation}
where we called:
\begin{align}
&\rho^{2}\equiv r^{2}+a^{2}\cos^{2}\theta ~,\label{rho1}\\ 
&\Sigma \equiv  r^{2}+a^{2}+Da^{2}\sin^{2}\theta~, \label{Sigma1} \\
& D\equiv \frac{R_S}{\rho^{2}}\left(r-\frac{\gamma^{2}\Delta R_{S}}{r^{2}}\right)~. \label{D1} 
\end{align}
The metric written in $\{ u,r,\theta,\phi\}$ coordinates has real-valued components and is axisymmetric; $a$ takes the physical meaning of the black hole spin parameter, related to the black hole angular momentum through $a=J/M$. This is the effective metric written in Kerr coordinates and is (coordinate) singularity free.
To better analyze the metric we can cast it in the Boyer-Lindquist form through the following coordinate transformation:
\begin{align}
 &u=t-\int \frac{r^{2}+a^{2}}{ r^{2}+a^{2}-rR_{S}+\frac{\gamma^{2}\Delta R_{S}^{2}}{r^{2}} } \dd r~,  \\
 &\phi=\psi-\int\frac{a}{r^{2}+a^{2}-rR_{S}+\frac{\gamma^{2}\Delta R_{S}^{2}}{r^{2}}}\dd r~.
\end{align}
The result in $ \{t,r,\theta,\psi\}$ coordinates is
\begin{equation}
g_{\mu \nu}=
\begin{pmatrix}
-1+D      & 0 & 0 & -a D\sin^{2}\theta  \\
    0    & \frac{\rho^{2}}{C} & 0 & 0 \\
   0 & 0 & \rho^{2} & 0 \\
-a D\sin^{2}\theta & 0 & 0& \Sigma \sin^{2}\theta 
\end{pmatrix} ~,
\label{BoLi}
\end{equation}
where we defined
\begin{equation}
    C\equiv r^{2}+a^{2}-\rho^{2}D~.
    \label{C1}
\end{equation}
The correct classical limit can be checked by sending $\Delta \rightarrow 0$, while for $a=0$ the metric reduces to \eqref{sch}.
\section{Main features of the effective Kerr metric}
\label{section4}
\subsection{ Domain and curvature invariants}
As already discussed, the spherically symmetric seed metric \eqref{PG} shows a lower bound in $r$ that is inherited by the effective axisymmetric metric since the coordinate $r$ is formally the same as in the spherically symmetric case:
\begin{equation}
r\geq (\gamma^2 \Delta R_S)^\frac{1}{3}\equiv r_{min}~.
\label{raggiominimo}
\end{equation}
We will give an interpretation to this lower bound when we compute the metric in generalized Painlevé-Gullstrand (GPG) coordinates.\par
The volume region within $r_{min}$ is not a sphere anymore but an oblate spheroid as is immediately clear looking at the relation between the Boyer-Lindquist and the Kerr-Schild coordinates:
\begin{equation}
    \begin{cases}
    &x^{2}+y^{2}=(r^{2}+a^{2})\sin^{2}\theta~, \\
    &z=r \cos\theta~.
    \end{cases}
\end{equation}
 This allows a generalization of some features of  the spherically symmetric seed metric, like the censorship of the ring singularity: such curvature singularity as we will show explicitly lies inside the forbidden region ($r=0$), and therefore does not belong to this spacetime. It is worth mentioning that the avoidance of the ring singularity is a common and expected feature of LQG-based effective Kerr metrics \cite{Muniz:2024wiv,Brahma:2020eos,Kumar:2022vfg,Suresh:2024hth,Islam:2022wck}. \par
It can be easily proved that as in the spherically symmetric case \cite{Kelly:2020uwj} the metric at $r_{min}$ is the Minkowski metric in oblate spheroidal coordinates, signaling that quantum repulsion there compensates exactly the gravitational attraction. This does not imply that curvature invariants are zero at $r=r_{min}$, since the metric is flat only on this surface.\par
The Ricci scalar for this metric reads:
\begin{equation}
R(r,\theta)=  -\frac{6 \Delta \gamma^{2}R_{S}^{2}}{r^{4}\rho^{2}} ~.
\end{equation}
In the equatorial plane $\theta=\pi/2$ it takes the same formal expression of the spherically symmetric one \cite{Kelly:2020uwj}, which can be recovered for any $\theta$ in the limiting case $a=0$. The divergence at $r=0$ (the classical ring singularity), is outside the domain of the metric tensor as discussed above. It 
has a global minimum $R(r_{min},\pi/2)=-\frac{6}{\gamma^2 \Delta}$ (the same value of the spherically symmetric one \cite{Kelly:2020uwj}) and rapidly decays to zero outside the deep quantum region. As in the nonrotating case, quantum gravity effects here can be seen as generating an effective stress-energy tensor, which in turn produces nonvanishing curvature invariants. 
\par
The Kretschmann scalar shows a divergence in $r=0$ as well, which is however outside the domain of the metric tensor. Its explicit expression is
\begin{align}
K=&\frac{36 R_{S}^{2}}{\rho^{12}r^{8}}\bigg[\gamma^{4}\Delta^{2}R_{S}^{2}\bigg( \cos^{8}\theta a^{8}+\frac{14 \cos^{6}\theta a^{6}r^{2}}{3}+\frac{26}{3}\cos^4\theta R_{S}^{2}a^{4}r^{4}+ \frac{22}{3}\cos^{2}\theta a^{2}r^{6}+13r^{8}\bigg)+\notag \\
&+\gamma^2 \Delta R_S \bigg({2}a^{6}\cos^{6}\theta +10 a^{4}r^{2}\cos^{4}\theta+{62}r^{4}a^{2}\cos^{2}\theta- {10}r^{6}   \bigg)\frac{ r^{5}}{3}+\notag \\
&+r^{8}\bigg(-\frac{\cos^{6}\theta a^{6}}{3}+5\cos^{4}\theta a^{4}r^{2}-5\cos^{2}\theta a^{2} r^{4}+\frac{r^{6}}{3}\bigg)\bigg] ~,
\end{align}
 showing a maximum at $(r_{min},~ \theta=\pi/2)$ and rapidly decaying to zero for $r>r_{min}$.

\subsection{Horizon structure and ergoregion}
\label{subsection4b}
The horizons can be found by imposing $1/g_{rr}=0$, which gives the loci where the $4$-acceleration of a static observer diverges. It reads
\begin{equation}
r^4+r^2a^2-R_{S}r^3+\gamma^2 \Delta R_{S}^2=0~.
\label{eventhorizon}
\end{equation}
As in the classical case, we can have sub-extremal black holes (two horizons), extremal black holes (only one horizon) and over-extremal black holes (no horizon at all).
 To distinguish between the various cases we can rearrange 
 \eqref{eventhorizon} as
\begin{equation}
 \frac{|a|}{R_S}=\sqrt{\frac{r}{R_S}\left(1-\frac{r}{R_S}\right)-\frac{\gamma^2\Delta}{{r}^2 }} ~; 
\end{equation}
the right-hand side has a global maximum:
\begin{equation}
\frac{|a_{max}|}{R_S}=\frac{1}{2}-\frac{2 \gamma^2 \Delta}{R_S^2} + O\left(\frac{\Delta^2}{R_S^4}\right)~,
\label{extremality}
\end{equation}
which corresponds to the spin for the effective extremal black hole. The location of the horizon in this case is given by
\begin{equation}
r^{\pm}_{H}|_{a_{max}}=\frac{R_{S}}{2}+\frac{8\gamma^2 \Delta }{R_{S}^2}+O\left(\frac{\Delta^{\frac{3}{2}}}{R_{S}^3}\right)~.    
\end{equation}
As in the classical case, for $a>a_{max}$ the interior becomes naked to an external observer, even if here it does not reveal a naked singularity. For $a<a_{max}$ we have two horizons, obtained expanding in powers of ${\Delta}/{R_{S}^2}$:
\begin{align}
&r^{-}_{H}=r^{-}_{H(cl.)} \left\{1- \frac{\gamma^2 \Delta R_{S}^2\left[r^{+}_{H(cl.)}(R_S^2-2a^2)-a^2 R_S \right]}{a^4\left( 4a^2-R_{S}^2  \right)r^{-}_{H(cl.)}  }  +O\left(\frac{\Delta^{\frac{3}{2}}}{R_{S}^3}\right)\right\}, \\
&r^{+}_{H}=r^{+}_{H(cl.)} \left\{1-\frac{\gamma^2 \Delta R_{S}^2\left[r^{-}_{H(cl.)}(R_S^2-2a^2)-a^2 R_S \right]}{a^4\left( 4a^2-R_{S}^2  \right)r^{+}_{H(cl.)}  } +O\left(\frac{\Delta^{\frac{3}{2}}}{R_{S}^3}\right)\right\} \label{outero},
\end{align}
where
\begin{equation}
r^{\pm}_{H(cl.)}=\frac{1}{2}\left( R_{S}\pm \sqrt{R_{S}^2-4a^2}\right)
\label{utero}
\end{equation}
are the classical locations of the horizons.
Differently from the spherically symmetric case where the effective metric acquires an inner horizon that is absent in the classical solution \cite{Kelly:2020uwj}, here the horizon structure remains almost unaltered and the location of the horizons acquires only a small quantum correction.
On the other hand, the ergoregion is affected significantly by quantum gravity. Let's call $\xi=\partial_t$ the asymptotic time-translation Killing vector;
the boundaries of the ergoregion can be found  by solving $\xi^\mu \xi_\mu=g_{tt}=0$:
\begin{equation}
r^4-r^3 R_{S}+r^2a^2\cos^2\theta +\gamma^2 \Delta R_{S}^2=0 ~.  
\label{ergo}
\end{equation}
We can study the solution by expanding it in powers of $\Delta/R_{S}^2$:
\begin{align}
&r_{E}^{-}=r_{E(cl.)}^{-}\left\{1+\frac{\gamma^2\Delta R_{S}^2\sec^2\theta\left[-a^2+R_{S}\sec^2\theta  ~r^{+}_{E(cl.)}    \right] }{a^4 r^{-}_{E(cl.)}\sqrt{R_{S}^2-4a^2\cos^2\theta}} +O\left(\frac{\Delta^{\frac{3}{2}}}{R_{S}^3}\right)\right\} \label{innerergo} ,\\
&r_{E}^{+}=r_{E(cl.)}^{+}\left\{1-\frac{\gamma^2\Delta R_{S}^2\sec^2\theta\left[-a^2+R_{S}\sec^2\theta  ~r^{-}_{E(cl.)}   \right] }{a^4r^+_{E(cl.)}\sqrt{R_{S}^2-4a^2\cos^2\theta}}+ O\left(\frac{\Delta^{\frac{3}{2}}}{R_{S}^3}\right)\right\},
\end{align}
where
\begin{equation}
 r_{E(cl.)}^{\pm}=\frac{1}{2}\left(R_{S}\pm \sqrt{R_{S}^2-4a^2\cos^2\theta   }\right)  .
\end{equation}
A remarkable curve in the ergoregion is the circumference given by the intersection of the inner ergosurface \eqref{innerergo} and the equatorial plane $\theta=\pi/2$, that classically coincides with the ring singularity. In the effective case, we can solve directly \eqref{ergo} for $\theta=\pi/2$, finding:
\begin{align}
r^{-}_{E}|_{\theta=\pi/2}=(\gamma^2 \Delta R_S)^{\frac{1}{3}}\left[1+\left(\frac{\gamma^2 \Delta}{3 R_S^2}\right)^\frac{1}{3}+O\left( \frac{\Delta}{R_S^2}\right) \right].
\label{ergopi}
\end{align}
For macroscopic black holes ($R_S/\sqrt{\Delta}\gg 1$) this means a significant shift of the inner ergosurface in the equatorial plane (notice that as in the classical case its location is independent of $a$), which is not surprising since the inner ergosurface, differently from the other surfaces analyzed before, lies in the deep quantum region. This shift of the inner ergosurface location reveals a very small region in the $\theta=\pi/2$ plane where particles are not forced to corotate with the black hole, since $r^-_{E}$ lies outside $r_{min}$. This correction suggests that quantum gravitational repulsion tends to slow down the frame dragging of Kerr spacetime. We will confirm this intuition in the next section, where we compute explicitly the angular velocity of a free falling observer. 
\subsection{Censorship of closed time-like curves}
The metric \eqref{BoLi} does not allow closed time-like curves to exist. This can be seen as a direct consequence of the minimum radius given by \eqref{raggiominimo}.If we consider a non-geodesic observer at $\theta={\pi}/{2},r=const.,t=const.$, the norm of its $4$-velocity in Boyer-Lindquist coordinates is
\begin{equation}
w^{\mu} w_{\mu}=g_{\psi \psi} \left(\frac{\dd \psi}{ \dd \lambda}\right)^{2}=\left[r^2+a^2+\frac{a^2 R_{S}}{r}\left(1-\frac{\gamma^{2}\Delta R_{S}}{r^3}\right)  \right]\left(\frac{\dd \psi}{ \dd \lambda}\right)^{2}~.
\end{equation}
Given the constraint \eqref{raggiominimo}, $r$ cannot be negative and the right-hand side of the previous expression is strictly positive; closed time-like curves are absent.\par
\begin{figure}
\centering
\includegraphics[scale=0.25]{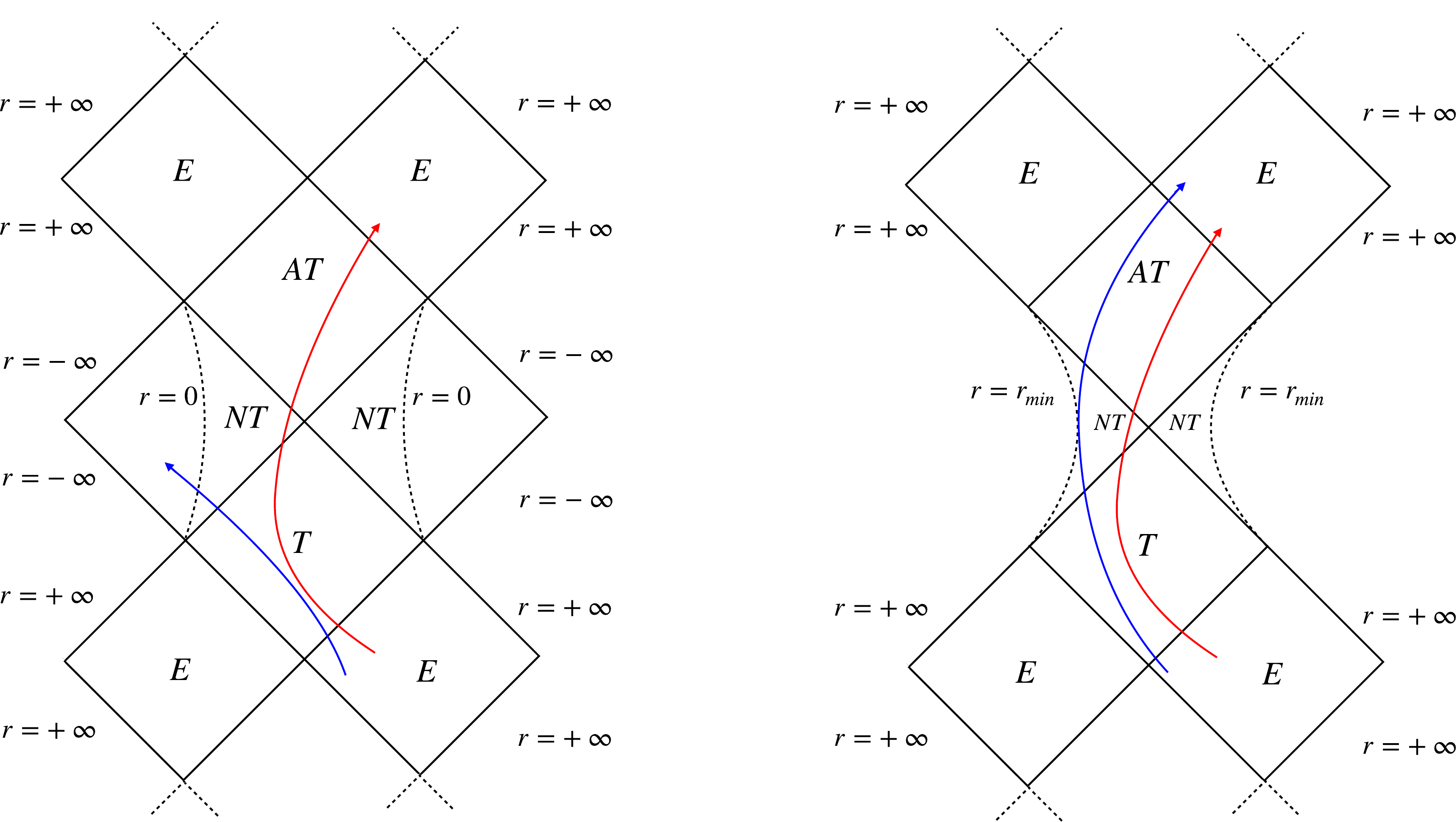}
    \caption[]{\footnotesize
    The plot shows the qualitative Penrose diagrams of the classical Kerr spacetime (left figure) and the effective one (right figure). The trajectories describe different journeys on the $\theta=0$ hypersurface. Here $E$ stands for external, $T$ for trapped, $AT$ for anti-trapped and $NT$ for non trapped regions.}    \label{figfazz}
\end{figure}

In the classical Kerr solution, a particle can travel across the disk $(r=0, \theta \neq \pi/2)$ inside the ring singularity ($r=0, \theta=\pi/2$) and emerge in the $r\leq 0$ submanifold, that has no horizons and is asymptotically flat (the blue curve in the left figure of Fig.~\ref{figfazz}). Closed time-like curves then are possible in the $\theta= \pi/2$ plane just outside the ring singularity in this portion of the manifold. In the effective spacetime presented here, the existence of a lower bound $r_{min}$ prohibits an observer from crossing the disk and entering in the $r<0$ region of the spacetime.
As we will see explicitly in the next section, a free-falling observer instead will stop his motion at $r_{min}$ and will revert it, as shown qualitatively in the right figure of Fig.~\ref{figfazz} (blue curve).\par
It is worth remarking that this does not exclude the possibility of finding the maximal extension of this spacetime. Recall indeed that in the maximal extension of the classical Kerr spacetime a free-falling observer can either enter the $r<0$ space through the equatorial disk and eventually reach its asymptotic region, or remain outside the ring and move toward the white hole inner and outer horizons. In the effective case only the second possibility is allowed.
A detailed analysis of the maximal extension of this spacetime is beyond the scope of this work and is left for future research.

\newpage
\section{Zero Angular Momentum Observer's journey}
\subsection{Quantum gravitational frame braking}
As in the spherically symmetric case, the inner boundary of the domain given by \eqref{raggiominimo} is manifest in the GPG form of the effective metric.\par 
To write the metric in such coordinates, we first need to derive the angular momentum and the proper radial velocity of the Zero Angular Momentum Observer (ZAMO), which generalizes the radially free-falling PG observer of the spherically symmetric solution. The ZAMO is an observer coming from infinity with zero angular momentum and is forced to corotate with the spacetime due to frame dragging. Its angular velocity $\Omega$ is defined as in the classical case:
\begin{equation}
 \Omega\equiv \frac{\dd \psi}{dt}=\frac{\dd \psi}{\dd \tau}\frac{\dd \tau}{\dd t}=\frac{u^{\psi}}{u^{t}}~,   \label{omega}
\end{equation}
where $u^{\psi}$ and $u^{t}$ are components of the observer's four-velocity.\par
Since by definition the observer at infinity has zero orbital angular momentum $L_{z}$ and ${\partial}/{\partial \psi}$ is a Killing vector, we have that
\begin{equation}
L_{z}=u_{\psi}=g_{\psi \psi}u^{\psi}+g_{\psi t}u^{t}=0 ~.   
\end{equation}
Therefore
\begin{equation}
 \Omega=-\frac{g_{\psi t}}{g_{\psi \psi}}=\frac{aD}{\Sigma}~,   \label{39}
\end{equation}
where $\Sigma$ and $D$ are given respectively by \eqref{Sigma1} and \eqref{D1}.
It is interesting to notice that $\Omega(r_{min},\theta)=0$ for any $\theta$; as already mentioned in section \ref{subsection4b} quantum gravity seems to be repulsive in the radial direction (as in the spherically symmetric case), but also in the angular direction, slowing down the spacetime frame dragging in the high curvature regime. We call this effect \emph{quantum gravitational frame braking}. \par
\begin{figure}
    \centering
    \includegraphics[scale=0.9]{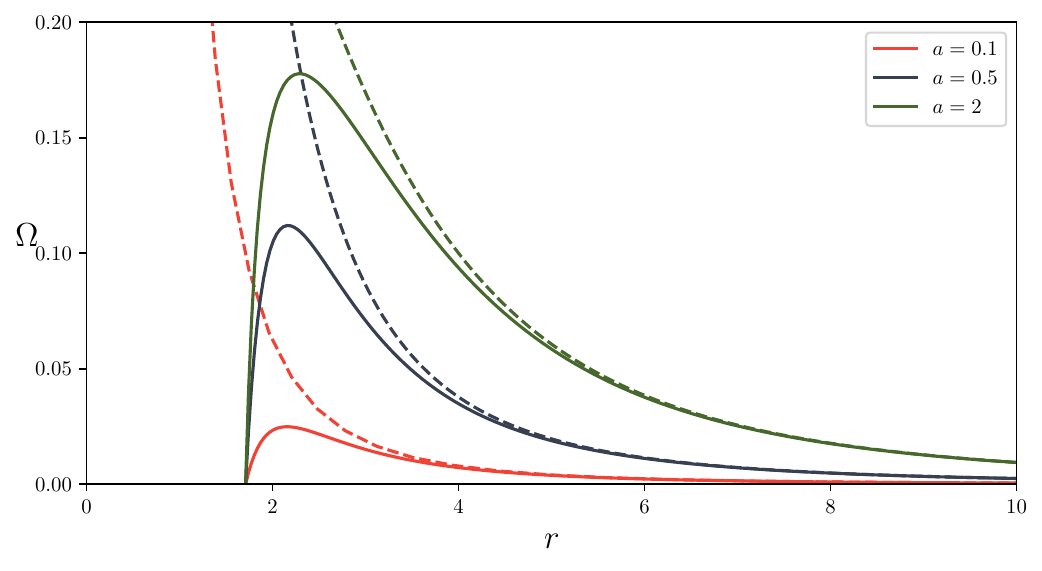}
    \caption[]{\footnotesize
    The plot shows the coordinate angular velocity $\Omega$ of a ZAMO for different values of $a$ for a journey on the equatorial plane $\theta=\pi/2$, both in the classical (dashed lines) and effective (solid lines) case. The classical curves diverge at $r=0$, while the effective ones have all an $a$-dependent maximum and a common zero at $r_{min}=(\gamma^2 \Delta R_S)^{\frac{1}{3}}$. For the plot we have set $R_S=5$, $\gamma=\Delta=1$. Notice that $a=2$ is sub-extremal in these units.}
    \label{figomega}
\end{figure}

We plot $\Omega(r, \theta=\pi/2)$ and compare it with its classical counterpart in Fig. \ref{figomega}. 
Two points are most relevant: the zero of the functions on $r_{min}$, as expected, and a global $a$-dependetent maximum for any curve. Imposing
\begin{equation}
\partial_r \Omega (r, \theta=\pi/2)=2 \gamma^2 \Delta R_{S}(2a^2+3r^2)-r^3(a^2+3r^2)=0 ~, 
\end{equation}
we find one positive real solution for the maximum at
\begin{equation}
r_{\Delta}=(4\gamma^2 \Delta R_S)^{\frac{1}{3}}\left[1-\left(\frac{2 \gamma^4 \Delta^2 R_S^2}{a^6} \right)^{\frac{1}{3}}+O\left(\frac{\Delta}{R_S^2} \right)^\frac{5}{3}\right].
\label{maximumrotation}
\end{equation}
As expected \eqref{maximumrotation} contains only terms involving $\Delta$ by the fact that is a pure quantum feature of the spacetime without classical counterpart. Moreover comparing \eqref{maximumrotation} with the location of the inner ergosurface at $\theta=\pi/2$  \eqref{ergopi} we see that the maximum lies inside the ergoregion, meaning that this quantum gravitational effect becomes relevant already in the ergoregion.

\subsection{Quantum gravitational radial braking}

Let's compute now the proper radial velocity $u^{r}={\dd r}/{\dd \tau}$ of the ZAMO. To perform the computation we use the energy conservation, the angular momentum conservation and the norm of the four velocity. After an immediate computation and using \eqref{39} we find
\begin{equation}
u^{r}=-\sqrt{\frac{D}{\rho^2}(r^2+a^2)}~.
\label{propervelzamo}
\end{equation}
The minus sign is chosen to require infalling motion. It can be checked that for $a=0$ this expression reduces to the radial proper velocity of an observer starting with zero velocity at infinity in the effective Schwarzschild metric \cite{Kelly:2020uwj}.\par
\begin{figure}
    \centering
    \includegraphics[scale=0.9]{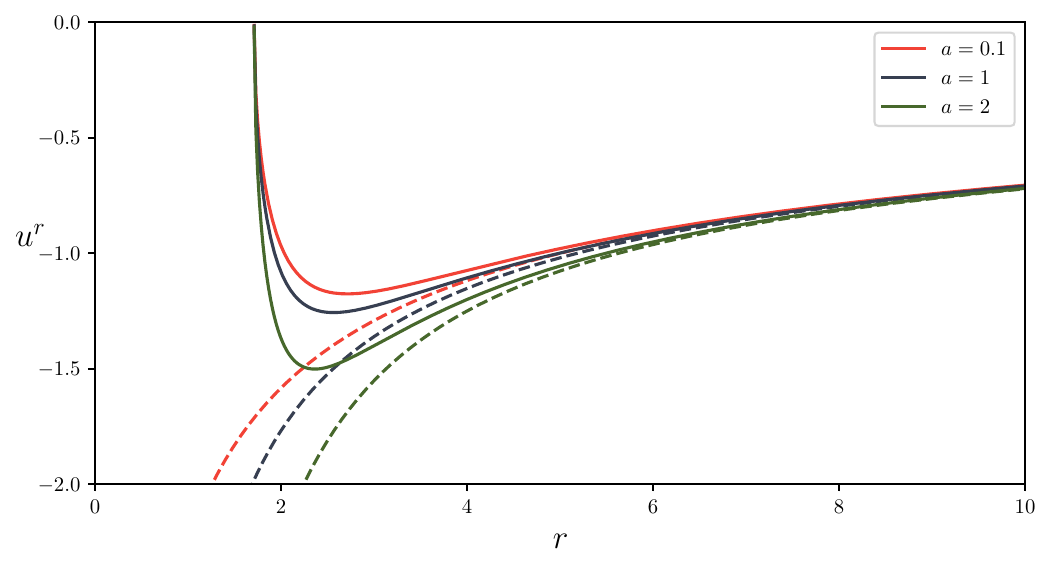}
    \caption[]{\footnotesize
    Radial proper velocity $u^r$ of a ZAMO for different values of $a$ for a journey on the equatorial plane $\theta=\pi/2$, in the classical (dashed lines) and effective (solid lines) case. The classical curves diverge at $r=0$, while the effective ones have all an $a$-dependent minimum and a common zero at $r_{min}=(\gamma^2 \Delta R_S)^{\frac{1}{3}}$. For the plot we have set $R_S=5$, $\gamma=\Delta=1$.}
    \label{figUR}
\end{figure}

 Fig.\ref{figUR} shows the radial proper velocity of the ZAMO  \eqref{propervelzamo} together with its classical counterpart.
While in the classical case the free falling observer will reach the ring singularity with an infinite proper radial velocity, here his motion stops at $r_{min}$ independently from the particular value of $a$, due to quantum radial repulsion of gravity in the high curvature regime. All the curves show a global minimum in the ergoregion. The ZAMO motion will continue outward with a radial velocity given by $u^r=\sqrt{\frac{D}{\rho^2}(r^2+a^2)}$, as schematically depicted in fig.~\ref{figfazz}.\par
By using these results we can already guess that regardless of the exact expression of the metric in GPG coordinates, the metric will not cover the whole range $r\in [0,+\infty)$ since the ZAMO cannot penetrate within $r_{min}=(\gamma^{2}\Delta R_{S})^\frac{1}{3}$. We will prove this statement in the next subsection.
\subsection{Generalized Painlevé-Gullstrand coordinates and minimum allowed radius}
We are now in the position to build the Painlevé-Gullstrand form of metric \eqref{Boli} (for the derivation in the classical case see \cite{Natario:2008ej}). The goal is to write the metric in the form ($\{T,r,\theta, \Phi \} $ coordinates):
\begin{align}
\dd s^{2}=&-\dd T^{2}+\gamma_{rr}(\dd r-v\dd T)^{2}+2\gamma_{r\theta}(\dd r-v\dd T)\dd \theta+\gamma_{\theta \theta}\dd \theta^{2}+ \notag \\ 
&+\Sigma \sin^2\theta(\dd \Phi-\omega \dd T)^{2}+2\gamma_{\theta \phi}\dd \theta (\dd \Phi-\omega \dd T)~, 
\label{metricPG}
\end{align}
where $v\equiv \dd r/  \dd T$ is the radial velocity of the ZAMO and $\omega\equiv \dd \Phi / \dd T$ is the angular velocity in the new coordinates. After an involved computation that is showed in Appendix \ref{appendix:PGcoord}, the effective metric in generalized Painlevé-Gullstrand coordinates reads:
\begin{align}
\dd s^{2}=&-\dd T^{2}+\frac{\rho^2}{\Sigma}(\dd r-v\dd T)^{2}+(\rho^{2}+\delta^{2}\sin^2\theta\Sigma) \dd \theta^2+ \notag \\
&+\Sigma \sin^2\theta(\dd \Phi-\omega \dd T)^{2}+2\Sigma \sin^2\theta ~\delta(r,\theta) \dd \theta (\dd \Phi-\omega \dd T) ,
\label{metricafinale}
\end{align}
where 
\begin{align}
   &v=u^r=-\sqrt{\frac{1}{\rho^2} (r^2+a^2)}~, \\
   & \omega=\Omega=\frac{a D}{\Sigma}~, \label{omegaO}\\
   & \delta(r,\theta)=a^{2}\sin(2\theta)\int \frac{v \omega}{\Sigma}\dd r~ . 
   \end{align}
It is worth mentioning that $v=u^r$ is expected since $T$ is the proper time of the ZAMO; the second equality ($\omega= \Omega$) is less obvious:  $\Omega$ is not a gauge invariant quantity and in general would transform under a coordinate transformation. For the explicit computation see \ref{appendix:PGcoord}.\par
The metric \eqref{metricafinale} is free from coordinate singularities as in the classical case, but differently from its classical counterpart is also free from physical singularities and is not defined everywhere. In particular, since it contains $v$, is defined only for 
$r\geq (\gamma^{2}\Delta R_{S})^{\frac{1}{3}}$.
As anticipated in the first section, this lower bound in the domain is not interpreted as a failure of PG coordinates in the region $r\in [0,r_{min})$: if we assume that such spacetime is generated by a collapsing rotating dust (or possibly rotating matter with pressure) source, the radius $r_{min}$ would be the minimum allowed radius for a collapsing star with total gravitational mass $M$. The quantum repulsive behaviour of gravity would not allow the star to contract further, as in the spherically symmetric case.\par
Both considerations have to be considered conjectures at this stage, since in order to be proved they require effective equations for axisymmetric spacetimes, a tool that is not yet available in literature.\par
To conclude this section we mention that casting an axisymmetric metric in a GPG form is a non-trivial result: it can be showed indeed that a different choice in the Janis-Newman algorithm does not admit a GPG form. For more details see Appendix \ref{appendix:otherPGcoordinates}

\section{Geodesics}
The explicit form of the geodesic equations for this spacetime is quite involved, but simplified by the non-trivial existence of an effective separation Carter constant. As in the classical case, it allows to decouple the equations and simplify their resolution. The derivation of the Carter constant and the geodesic equations is given in Appendix \ref{appendix:cartekconstant}. Here we will specialize the analysis to equatorial time-like and null circular geodesics.

\subsection{Equatorial time-like circular geodesics and modification of the third Kepler law}

We look at the equatorial ($\theta=\pi/2$) circular motion  of a massive particle in this background, seeking for a modification of the third Kepler law; here we will work in Boyer-Lindquist coordinates.
From the geodesic lagrangian
\begin{equation}
\mathcal{L}=\frac{1}{2}g_{\mu \nu} \Dot{x}^{\mu} \Dot{x}^{\nu}
\end{equation}
we can derive the geodesic trajectory from the Euler-Lagrange equations:
\begin{equation}
 \frac{\dd}{\dd\lambda}\left(\frac{\partial \mathcal{L}}{\partial \Dot{x}^{\mu}}\right)=\frac{\partial \mathcal{L}}{\partial x^{\mu}}~.
 \label{EL}
\end{equation}
The $t,\theta$ and $\psi-$components at $\theta=\frac{\pi}{2}$ give respectively: 
\begin{equation}
 \quad u_{t}=-E~,\quad \Dot{\theta}=0~, \quad u_{\psi}=L ~, 
\end{equation}
as expected. By imposing $\dd^2{r}/\dd \tau^2=0$, $\dd r / \dd \tau=0$ to the remaining component, after a brief computation we find:
\begin{equation}
\partial_{r}D-2a\omega \partial_{r}D+\omega^2  \left({2r}+a^2\partial_{r}D \right)=0~,  
\label{angularvel1}
\end{equation}
where $\omega\equiv (\dd \psi/ \dd \tau)/(\dd t / \dd\tau) $ and should not be confused with the proper angular velocity of the ZAMO. Notice that this result holds only for orbits in the equatorial plane.\par
Since \eqref{angularvel1} is a quadratic equation in $\omega$, it is straigthforwardly solved by:
\begin{equation}
\omega_{\pm}=\frac{\frac{a R_S }{r^2}\left(\frac{4\gamma^{2}\Delta R_{S}}{r^3}-1 \right)\pm \sqrt{\frac{2R_S}{r}\left(1-\frac{4 \gamma^2 \Delta R_{S}}{r^3} \right)    }     }{2r+\frac{a^2 R_S}{r^2}\left(\frac{4\gamma^{2}\Delta R_{S}}{r^3}-1 \right)}
\label{newomega}
\end{equation}
which is the \emph{modified third Kepler law}.
 This coordinate angular velocity holds both for time-like, null and space-like geodesics.\par
\begin{figure}
    \centering
    \includegraphics[scale=0.9]{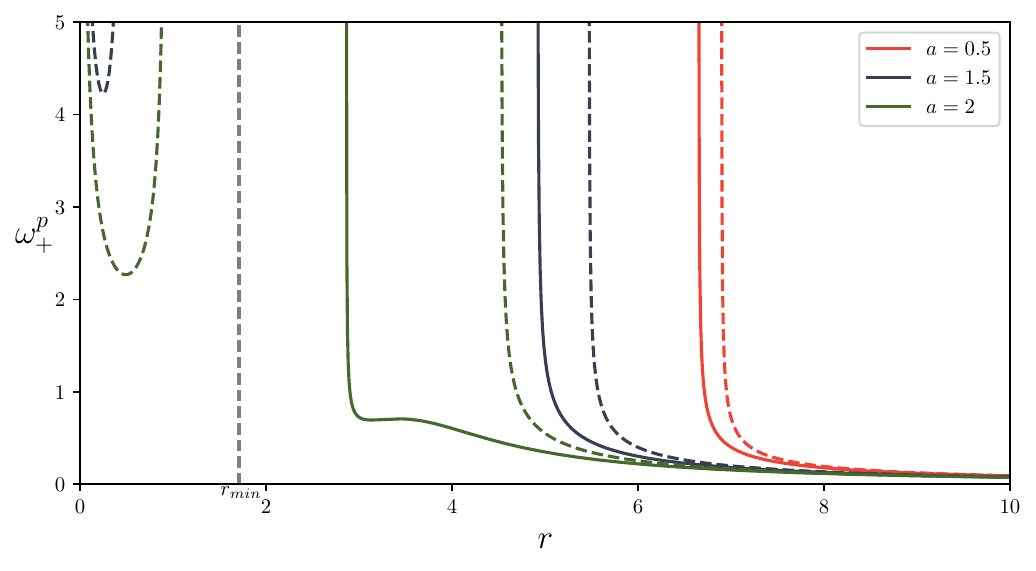}
    \caption[]{\footnotesize
    The plot shows the proper angular velocity  $\omega_+^p$ as a function of $r$ for a circular time-like geodesic
    on the equatorial plane $\theta=\pi/2$, both in the classical (dashed lines) and effective (solid lines) case, for $R_S=5$. For the plot we have set $\gamma=\Delta=1$.}
    \label{figOR}
\end{figure}

 To find the proper angular velocity for time-like geodesics we need to impose the normalization of the four-velocity.
 After a brief computation we get:
\begin{equation}
 \omega^P_{\pm}=\frac{\omega_{\pm}}{\sqrt{1-D+2aD\omega_{\pm}-\Sigma(\omega_{\pm})^2 }} ~,
\end{equation}
where we called $\omega^P$ the proper angular velocity and $\omega_{\pm}$ is given by \eqref{newomega}.
Fig.\ref{figOR} shows the function $\omega^P_+$ compared with its classical counterpart.
For small $a$ the behaviour is similar to the classical one, and in particular $\omega_+^p$ is a monotonic decreasing function of $r$. Close to the extremal case \eqref{extremality} instead the plot shows a local maximum and minimum (green solid line).\par 
All the effective curves tend to the Kepler law $\omega= \sqrt{{R_S}/{(2r^3)}}$ for large $r$, independently from the particular value of $a$. Also, both the classical and effective curves diverge near the location of the outer horizons, given respectively by \eqref{outero} and \eqref{utero}.\par
An interesting feature of the effective angular velocity is that is not defined for $r\geq (4\gamma^2 \Delta R_S)^\frac{1}{3}$ (the same holds for the non-rotating case $a=0$), as one can easily notice from \eqref{newomega}; this means that in the non-trapped region $r_{min}<r<r_-^H$ equatorial circular time-like geodesics are forbidden (non-geodesic circular trajectories could in principle be allowed). This feature is interpreted by the Author as an effect of quantum gravity (such region lies in the deep quantum gravity regime), which produces a physical repulsion. This is opposite to what happens in the non-trapped inner region of the classical Kerr metric, where equatorial circular time-like geodesics are allowed for $0<r<r_-^H$ due to the attractive nature of classical gravity, as seen in Fig.\ref{figOR}. We conclude by remarking that the departure from the classical case of the effective curves are magnified by the choice $\Delta=1$ for the plot. 

\subsection{Effective potential for null equatorial geodesics and photon sphere}
Since computations for geodesics in the effective Kerr spacetime are difficult to be handled, to study the photon equatorial trajectories is useful to look at the null effective potential; the computations are formally equivalent to the classical case \cite{Ferrari:2020nzo}.
From the normalization of the four-velocity and the conserved energy and angular momentum, it can be easily showed that for equatorial photons holds:
\begin{equation}
 \left(\frac{\dd r}{\dd \lambda} \right)^2=\frac{\Sigma}{r^2}(E-V_+)(E-V_-)  ~,
 \label{epiumeno}
\end{equation}
where
\begin{equation}
V_{\pm}=\frac{L}{\Sigma}(aD\pm \sqrt{C})  
\label{Vpm}
\end{equation}
and $\Sigma$, $D$, $C$ are evaluated at $\theta=\pi/2$. $E$ and $L$ are the conserved Killing quantities for equatorial photons. It is clear from \eqref{Vpm} that $V_{\pm}$ depend both on $a$ and $L$; we have therefore four possible cases: $(a\lessgtr 0,L\lessgtr 0)$. It can be easily showed that the combinations for which $aL>0$ have the same $V_\pm$ (similarly for $aL<0$), simply switched in names; therefore only two relevant cases are left: the corotating case ($aL>0$) and the counter rotating one ($aL<0$), independently from the particular direction of the rotation. Let's analyze the case $aL>0$. Following \cite{Ferrari:2020nzo}, to simplify the notation we redefine $V_{\pm}$ in the following way:
\begin{align}
 & V_{-}=\frac{1}{\Sigma}(LaD-|L|\sqrt{C}) ~, \\
 &V_{+}=\frac{1}{\Sigma}(LaD+|L|\sqrt{C}) ~. 
 \end{align}
In this way for both cases $V_-<V_+$. We firstly notice that $V_+=V_-$ only for $C=0$, that gives the location of the two horizons $r^{+}_{H}$ and $r^{-}_{H}$, similarly to the classical case. For $r^{-}_{H}<r<r^{+}_{H}$ the potential is not defined, while for the remaining range $V_{-}<V_{+}$. As in the classical case, it can be easily checked that for $r\rightarrow +\infty$, $V_{\pm}\rightarrow 0$. In the $aL>0$ case $V_+$ is positive in its whole domain of definition, while  $V_{-}$ has zeroes for:
\begin{equation}
aD=\sqrt{C} ~.    
\end{equation}
This equation is satisfied only for $D=1$, that means
\begin{equation}
  r^4-r^3R_S+\gamma^2 \Delta R_S^2=0  ~.
\end{equation}
By comparing the previous expression with \eqref{ergo} is clear that the locations of the zeroes of $V_{-}$ are the ergosurfaces $r^{\pm}_{E}|_{\theta=\pi/2}$ as in the classical case.\par
Finally, we can study the stationary points of $V_{+}$ and $V_{-}$. $V_{+}$ has a maximum for $r_{H}^{+}<r<r_{E}^+$ and $V_{-}$ a minimum for $r>r_E^+ $.\par
\begin{figure}
    \centering
    \includegraphics[width=0.9\textwidth]{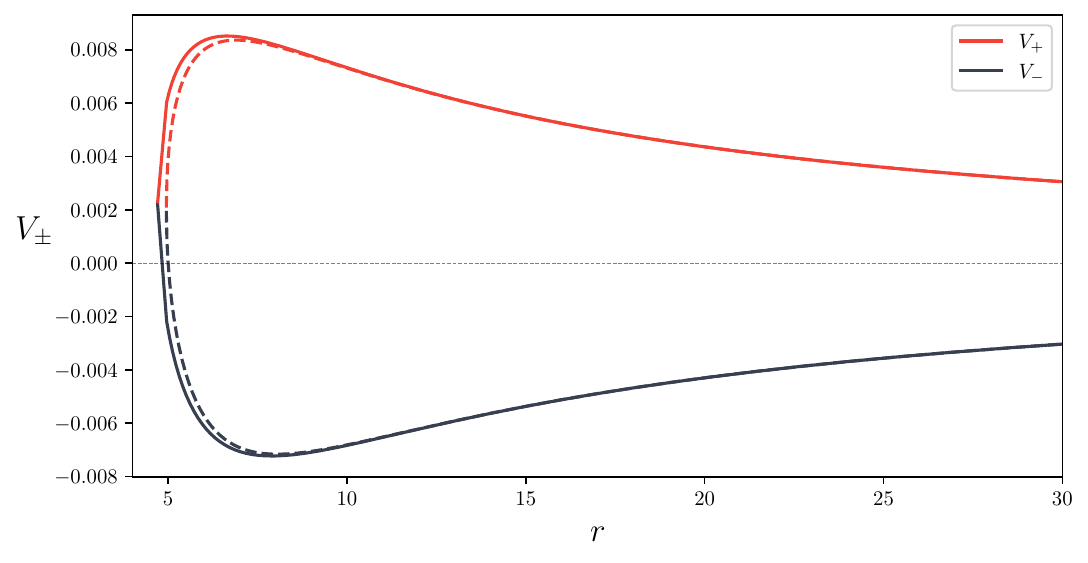}
    \caption[]{Shape of $V_{+}$ and $V_{-}$ for co-rotating equatorial photons in the classical (dashed lines) and effective (solid line) cases, for $r>r_{min}=(\gamma^2 \Delta R_{S})^{\frac{1}{3}}$. The parameters have been fixed to $R_S=5,$ $a=0.5$, $\Delta=1$, $\gamma=1$.}
    \label{fig:corotating}
\end{figure}

The two functions are shown in Fig.\ref{fig:corotating}. From the plot it is clear that they are qualitatively equivalent to their classical counterparts, with small departures close to the outer horizon $r^{+}_H$.
The shape of the potentials for the counter rotating case ($aL<0$) is the same, reflected about to the $r$-axis.\par
Let's look now at the equatorial photon sphere  ($\dd{r}/ \dd \lambda=\dd ^2r/ \dd \lambda^2=0$) in the case $aL>0$. In order to find the equatorial photon circle, we can differentiate \eqref{epiumeno} with respect to the affine parameter:
\begin{equation}
\frac{\dd ^2 r}{\dd \lambda^2}=\frac{1}{2}\partial_{r}\left(\frac{\Sigma}{r^2}\right)(E-V_{+})(E-V_{-})-\frac{\Sigma}{2r^2}\left[(E-V_{-})\partial_{r}V_{+}+(E-V_+)\partial_{r}V_{-}  \right]   ~. 
\end{equation}
Now, since we are interested in circular orbits, we can impose $\dd r / \dd \lambda=0$, and using \eqref{epiumeno} we have:
\begin{align}
&\frac{\dd ^2 r}{\dd \lambda^2}=-\frac{\Sigma}{2r^2}(V_{+}-V_{-})\partial_{r}V_{+}, \quad for ~~ E=V_{+}, \\
& \frac{\dd ^2 r}{\dd \lambda^2}=\frac{\Sigma}{2r^2}(V_{+}-V_{-})\partial_{r}V_{-}~, \quad for ~~ E=V_{-} ~.
\end{align}
Now, since:
\begin{equation}
 V_+-V_-=\frac{2|L|\sqrt{C}}{\Sigma} ,  
\end{equation}
then:
\begin{equation}
\frac{\dd ^2 r}{\dd \lambda^2}=\mp\frac{|L|\sqrt{C}\partial_{r}V_{\pm}}{r^2}~,\quad for ~~ E=V_{\pm}  ~.
\end{equation}
By imposing ${\dd ^2 r}/{\dd \lambda^2}=0$, we have four possibilities for corotating orbits: $L=0$, $C=0$, $\partial_{r}V_+=0$, $\partial_{r}V_-=0$. For $L=0$, in order to have $\dot{r}=0$ we should require $\Sigma=0$ (from \eqref{epiumeno}), and one can easily check that the resulting equation has no solutions; on the other hand $C=0$ coincides with the location of the horizons. The condition $\partial_r V_{+}=0$ gives the following:
\begin{equation}
36 \gamma^4 \Delta^2 R_{S}^4+8\gamma^2\Delta R_{S}^2(4 a^2 r^2-\frac{9}{2}R_{S} r^3+3 r^4)+r^5(-12R_{S} a^2   +9 R_{S}^2 r-12R_{S}r^2+4r^3)=0
\end{equation}
This equation can be solved numerically, giving two unstable circular orbits forming the effective photon sphere, one for $aL>0$ and the other $aL<0$, very close to the classical ones. Corotating photons coming from $r=+\infty$ with $E>V_{+}$ will reach $r_{min}$ and enter the matter-filled region (in the classical case they will hit the ring singularity at $r=0$); photons with $0<E<V_{+}$ will be deflected by the potential and photons with negative energies cannot reach (or come from) infinity (the case we are interested in). It can be proved that counter rotating photons can have negative energy in the ergoregion as in the classical case, and this allows the Penrose process. A detailed analysis of the variation in efficiency of the Penrose process for this effective space-time is outside the aims of this work and will be analyzed in a future work.

\section{Summary and conclusion}
\label{s.con}
We have studied an effective Kerr metric with quantum corrections inspired by loop quantum gravity. The metric has been derived through the Newman-Janis algorithm applied to a spherically symmetric LQG-based metric widely studied in literature. The main features of this effective axisymmetric metric are: the absence of the ring singularity, the absence of closed time-like curves and a minimum radius $r_{min}=(\gamma^2 \Delta R_S)^{\frac{1}{3}}$ inherited by its spherically symmetric counterpart. The repulsive behaviour of quantum gravity produces two novel effects that have no classical counterpart, analyzed through the motion of a ZAMO: the quantum gravitational radial braking that forces the observer to stop his motion at $r_{min}$, and the quantum gravitational frame braking that counteracts the classical frame dragging of Kerr spacetime, decreasing the ZAMO's angular momentum in the deep quantum region and countering it at $r_{min}$. The horizon structure of this spacetime is qualitatively equivalent to the classical one, while the inner boundary of the ergoregion is shifted significantly from its classical location at $r=0$. This is consistent with the quantum frame braking effect. \par
We derived the GPG form of the metric, which turns out to be naturally adapted to the lower bound $r_{min}$ as in the spherically symmetric case; a discussion about the nature and the physical interpretation of this lower bound is given.\par
We constructed the geodesic equations through the help of an effective Carter constant from the Hamilton-Jacobi approach, and solutions are studied for equatorial time-like and null circular orbits. We derived an effective third Kepler law valid for any kind of equatorial circular geodesics, and then we specialized it to time-like geodesics, showing that the inner non-trapped region forbids time-like circular geodesic motion due to the repulsive behaviour of quantum gravity. Then we derived an effective potential for null geodesics and the defining equation for the location of the photon ring.  \par
Further works in this direction will be the derivation of the maximal extension of this spacetime, its thermodynamical properties, possible astrophysical signatures like quasi-normal modes and, more ambitiously, checking if axisymmetric effective equations could admit this metric as vacuum solution.

\appendix

\section{Metric in generalized Painlevé-Gullstrand coordinates}
\label{appendix:PGcoord}
In this appendix we derive the metric in GPG coordinates; the computation strictly follows the classical one \cite{Natario:2008ej}. We start from the effective metric in Boyer-Lindquist coordinates $\{t,r,\theta,\psi\}$:
\begin{equation}
 \dd s^{2}=-(1-D)\dd t^{2}-2a D \sin^2\theta\dd t \dd \psi+\frac{\rho^{2}}{C}\dd r^{2}+\rho^{2}\dd \theta^{2}+\sin^2\theta\Sigma \dd \psi^{2} .
 \label{metricBL}
\end{equation}
$\rho, \Sigma,D, C$ are defined respectively by \eqref{rho1},\eqref{Sigma1}, \eqref{D1},\eqref{C1}. Let us perform the following coordinate transformation
\begin{equation}
    \begin{cases}
&\dd t=\dd T+\alpha(r,\theta)\dd r+\gamma(r,\theta)\dd \theta ~,\\
& \dd \psi=\dd \Phi+\beta(r,\theta)\dd r+\delta(r,\theta)\dd \theta   ~,  \label{syst}
    \end{cases}
\end{equation}
where $\{T,r,\theta, \Phi \}$ are the new coordinates. If we plug this transformation inside the metric \eqref{metricBL} we get:
\begin{align}
 &\dd s^{2}=-(1-D)(\dd T+\alpha \dd r +\gamma \dd \theta)^{2}-2a D\sin^2\theta(\dd T+\alpha \dd r +\gamma \dd \theta)(\dd \Phi+\beta \dd r +\delta \dd \theta)+ \notag \\
 &+\frac{\rho^{2}}{C}\dd r^{2}+\rho^{2}\dd \theta^{2}+\Sigma\sin^2\theta(\dd \Phi+\beta \dd r +\delta \dd \theta)^{2}.  \label{PG2} 
\end{align}
Now we require that the term $g_{r\Phi}$ vanishes, as expected by the GPG form of the metric \eqref{metricPG}. This leads to
\begin{equation}
 \beta=\frac{aD}{\Sigma}\alpha=\Omega \alpha ~.   \label{beta1}
\end{equation}
Notice that $g_{r\Phi}=0$ means requiring that in the new coordinates the angular momentum of the ZAMO is yet given by \eqref{39}. Indeed, the angular momentum in the new coordinates is given by
\begin{equation}
L=u_{\Phi}=g_{\Phi\Phi}u^{\Phi}+g_{\Phi T}u^{T}+g_{\Phi r}u^{r}=g_{\Phi\Phi}u^{\Phi}+g_{\Phi T}u^{T}~,
\end{equation}
and since by definition of ZAMO $L=0$, we have:
\begin{equation}
 \omega\equiv \frac{\dot \Phi}{\dot T}=\frac{aD}{\Sigma}=\Omega~,   
\end{equation}
where we denote with dot the derivative with respect to the proper time of the ZAMO. 
The last equality is not obvious since $\Omega$ is not a scalar; once we have the metric in GPG coordinates we will prove explicitly that the coordinate transformation doesn't affect $\omega$. \par
Following \cite{Natario:2008ej} we further assume that in the new coordinates the radial velocity of the ZAMO is yet given by \eqref{39} (this is reasonable, since the proper time and the GPG time should coincide). Therefore
\begin{equation}
 v\equiv \frac{\dd r}{\dd T}=-\sqrt{\frac{1}{\rho^{2}}(\Sigma-C)}=-\sqrt{\frac{1}{\rho^{2}}D(r^{2}+a^{2})}~.
\end{equation}
Notice that the radial velocity of the ZAMO depends on $\theta$ through $\rho$.\par
These are the quantities that should enter in the GPG form of the metric \eqref{metricPG}. Let's assume this, and see if we reach an absurd. We proceed by imposing the equality between the $\dd T^{2}$ terms in the metrics \eqref{metricPG} and \eqref{PG2}. After a brief computation, we obtain:
\begin{equation}
\gamma_{rr}=\frac{\rho^{2}}{\Sigma}~.   \label{grr}
\end{equation}
By equating the coefficients of $\dd t \dd r$, we get:
\begin{equation}
\alpha=\frac{v \rho^{2}}{C}=-\frac{1}{C}\sqrt{\rho^{2}D(r^{2}+a^{2})} ~.
\end{equation}
Since neither $C$ nor $\rho^{2}D$ contain $\theta$, we have the important result that $\alpha=\alpha(r)$. This brings to a great simplification, since $\gamma$ in \eqref{syst} has to depend only on $\theta$; if indeed $\gamma$ depends on $r$, the transformation rule should have this form:
\begin{equation}
 t=T+f(r,\theta)  ~, 
\end{equation}
and could not be \eqref{syst}.
We can require that $\gamma|_{r=+\infty}=0$, which imposes $\gamma=0$ (since $\gamma$ is independent from $r$). Given \eqref{beta1}, we can already compute $\delta$. From the second transformation in \eqref{syst} we have to impose:
\begin{equation}
 \partial_{r}\int \delta(r,\theta) \dd \theta =\beta=\omega \alpha  ~.
\end{equation}
Integrating both sides over $r$, and deriving in $\theta$, we get:
\begin{equation}
\delta(r,\theta)=a^{2}\sin(2\theta)\int \frac{v \omega}{\Sigma}\dd r ~. 
\label{delta}
\end{equation}
We don't try to compute \eqref{delta} since is outside the aims of this work. \par
 By equating the $\dd r^{2}$ terms, we find after a bit of computations:
\begin{equation}
 C=r^2+a^2-D\rho^2 ~,   
\end{equation}
which leads to no other constraints, but also no contradictions. By comparing the terms $\dd r \dd \theta$ we find instead {$\gamma_{r \theta}=0$}. The terms $\dd t \dd \Phi$ are automatically equal.\par
By comparing the terms $\dd \theta^{2}$ we find:
\begin{equation}
\gamma_{\theta \theta}=\rho^{2}+\delta^{2}\Sigma \sin^2\theta~.  
\end{equation}
Finally, the comparison of the $\dd T \dd \theta$ terms gives:
\begin{equation}
\gamma_{\theta \Phi}=\delta~ \Sigma \sin^2\theta ~.   
\end{equation}
This concludes the matching of the components, and we did not meet any absurd from the assumption on $v$. Finally, the metric reads:
 \begin{align}
\dd s^{2}=&-\dd T^{2}+\frac{\rho^2}{\Sigma}(\dd r-v\dd T)^{2}+(\rho^{2}+\delta^{2}\Sigma\sin^2\theta) \dd \theta^{2}+ \notag \\
&+\Sigma \sin^2\theta(\dd \Phi-\Omega \dd T)^{2}+2\Sigma ~\delta\sin^2\theta \dd \theta (\dd \Phi-\Omega \dd T) ~.
\label{finalmetricPG}
\end{align}
where $\delta$ is given by \eqref{delta}.
The coordinate transformation that casts \eqref{metricBL} into \eqref{finalmetricPG} is:
\begin{equation}
    \begin{cases}
&\dd t=\dd T+\frac{v \rho^2}{C}\dd r  ~,\\
& \dd \psi=\dd \Phi+\Omega\frac{v \rho^2}{C}\dd r+\delta(r,\theta)\dd \theta   ~.  \label{syst1}
    \end{cases}
\end{equation}
In conclusion, we show explicitly the non-trivial result:
\begin{equation}
 \Omega\equiv \frac{\dot \psi}{\dot t}=\frac{\dot \Phi}{\dot T}\equiv \omega   
\end{equation}
using the transformation \eqref{syst1}: 
\begin{align}
 &\frac{\dd \psi}{\dd \tau}=\frac{\partial \psi}{\partial \Phi}\frac{\dd \Phi}{\dd \tau}+\frac{\partial \psi}{\partial r}\frac{\dd r}{\dd \tau}=\frac{\dd \Phi}{\dd \tau}+\frac{\Omega v \rho^2}{C}\frac{\dd r}{\dd \tau} ~, \\
 &\\
 &\frac{\dd t}{\dd \tau}=\frac{\partial t}{\partial T}\frac{\dd T}{\dd \tau}+\frac{\partial t}{\partial r}\frac{\dd r}{\dd \tau}=\frac{\dd T}{\dd \tau}+\frac{ v \rho^2}{C}\frac{\dd r}{\dd \tau}
 \end{align}
and recalling that $\Omega={\dot \psi}/{\dot t}$ we get:
\begin{equation}
 \frac{\dot \Phi}{\dot T}= \Omega   ~.
\end{equation}

\section{A different effective axisymmetric metric with no GPG form}
\label{appendix:otherPGcoordinates}
By applying the Newman-Janis algorithm to the seed metric \eqref{sch} we have in principle the freedom in the choice of the transformation \eqref{newz}, with no physically (or mathematically) motivated prescriptions, at least until effective equations of motion become available.
A different possible choice for the transformation is given by:
\begin{align}
 &l^{\mu}\rightarrow \Tilde{l}^{\mu}=\delta^{\mu}_{1}~,   \\
&n^{\mu} \rightarrow \Tilde{n}^{\mu}=\delta^{\mu}_{0}-\frac{1}{2}\left[1-\frac{R_S}{2}\left(\frac{1}{\Tilde{r}}+\frac{1}{\Tilde{\Bar{r}}} \right)+\frac{\gamma^{2}\Delta R_{S}^{2}}{(\Tilde{r}\Tilde{\Bar{r}})^2} \right]\delta^{\mu}_{1} ~, \\
& m^{\mu}\rightarrow \Tilde{m}^{\mu}=\frac{1}{\sqrt{2}\Tilde{r}}\left( \delta^{\mu}_{2}+\frac{i}{\sin{\Tilde{\theta}}}\delta^{\mu}_{3}\right)~.
\end{align}
It produces the following line element in Boyer-Lindquist coordinates:
\begin{equation}
g_{\mu \nu}=
\begin{pmatrix}
-1+\tilde{D}      & 0 & 0 & -a \tilde{D}\sin^{2}\theta  \\
    0    & \frac{\rho^{2}}{\tilde{C}} & 0 & 0 \\
   0 & 0 & \rho^{2} & 0 \\
-a \tilde{D}\sin^{2}\theta & 0 & 0& \tilde{\Sigma} \sin^{2}\theta 
\end{pmatrix} ~,
\label{BoLi2}
\end{equation}
where:
\begin{align}
 &\tilde{D}\equiv \frac{R_S }{\rho^2}\left(r-\frac{\gamma^2 \Delta R_S}{\rho^2} \right) ~, \\
 & \tilde{C}\equiv r^2+a^2-\rho^2 \tilde{D}  ,\\
 & \tilde{\Sigma}\equiv r^2+a^2+a^2\tilde{D} \sin^2\theta  
\end{align}
and $\rho^2=r^2+a^2\cos^2\theta$. Following the steps as in Appendix \eqref{appendix:PGcoord}, it can be showed that such line element cannot be cast in the form \eqref{metricPG}. 

\section{Geodesic equations and Carter constant}
\label{appendix:cartekconstant}
We derive here the geodesic equations for generic (also non-equatorial) trajectories. This can be quite easily done within the Hamilton-Jacobi approach, as in the classical case (see e.g.\cite{Ferrari:2020nzo}).
We start from the particle lagrangian:
\begin{equation}
L(x^{\mu},\Dot{x}^{\mu})=\frac{1}{2}g_{\mu \nu} \Dot{x}^{\mu} \Dot{x}^{\nu} ~.    
\end{equation}
where dot is differentiation with respect to the parameter $\lambda$. We define the conjugate momenta:
\begin{equation}
 p_{\mu}\equiv\frac{\partial L}{\partial \Dot{x}^{\mu}}=g_{\mu \nu}\Dot{x}^{\nu}  ~; 
\end{equation}
by inverting the previous we have $\Dot{x}^{\mu}=g^{\mu \nu} p_{\nu}$. Then we define the hamiltonian as:
\begin{equation}
 H(x^{\mu}(p_{\nu}),p_{\nu})=p_{\mu}\Dot{x}^{\mu}(p_{\nu})-L(x^{\mu},\Dot{x}^{\mu}(p_{\nu}))  =\frac{1}{2}g^{\mu \nu}p_{\mu}p_{\nu} 
\end{equation}
and we call:
\begin{equation}
  S\equiv S\left(x^{\mu},\lambda\right)  
\end{equation}
the solution of the Hamilton-Jacobi equation:
\begin{equation}
H\left(x^{\mu},\frac{\partial S}{\partial x^{\mu}}\right)+\frac{\partial S}{\partial \lambda}=0  ~,
\label{HJ}
\end{equation}
where the solution $S$ is the on-shell action. It can be proved that if $S$ is solution of the previous first order partial differential equation it 
 has to depend on 4 integration constants
 and has to satisfy:
\begin{equation}
 \frac{\partial S}{\partial x^{\mu}}=p_{\mu} ~.
 \label{conditionS}
\end{equation}
Let's solve \eqref{HJ}. A geodesic particle (massive or massless) moving in this effective spacetime has two conserved quantities: $p_{0}=-E$ (due to stationareity) and $p_{\psi}=L$ (due to axis-symmetry). Therefore, we have three equations:
\begin{equation}
  \begin{cases}
   & H=\frac{1}{2}g^{\mu \nu}p_{\mu}p_{\nu}=\frac{1}{2}k  ~,   \\
   &  p_{0}=-E  ~,  \\
   & p_{\psi}=L ~,
  \end{cases}  
  \label{sistema}
\end{equation}
where $k=-1$ for time-like geodesics, $k=0$ for null and $k=1$ for space-like geodesics. Now, to be $S$ solution of \eqref{HJ}, because of \eqref{conditionS}) we need:
\begin{equation}
 \frac{\partial S}{\partial t}=-E, \quad \frac{\partial S}{\partial \psi}=L.   
\end{equation}
Plugging the first of \eqref{sistema} in \eqref{HJ} we get
\begin{equation}
 \frac{\partial S}{\partial \lambda}=-\frac{1}{2}k   ~.
\end{equation}
We can therefore write the solution $S$ as
\begin{equation}
 S=-Et+L\psi-\frac{1}{2}k \lambda +S(r,\theta) ~.   
 \label{action}
\end{equation}
Now, as in the classical case \cite{Ferrari:2020nzo}  we make the ansatz of separability of $S$:
\begin{equation}
 S(r,\theta)=S^{r}(r)+S^{\theta}(\theta)~.  \label{solS}
\end{equation}
We expect that $S$ contains the fourth integration constant. By plugging \eqref{solS} in \eqref{HJ} and using the inverse of the effective metric \eqref{BoLi} we get
\begin{equation}
 -E^{2}\frac{\Sigma}{C}+\left( \frac{\partial S^r}{\partial r}\right)^2 \frac{C}{\rho^2}+\left( \frac{\partial S^\theta}{\partial \theta}\right)^2\frac{1}{\rho^2}+2EL\frac{aD}{C}+L^2\frac{1-D}{C \sin^2\theta}-k=0~.
 \end{equation}
By multiplying both members by $\rho^2$, and using the following identities:
\begin{align}
& \frac{\rho^2 \Sigma}{C}=\frac{(r^2+a^2)^2}{C}-a^2 \sin^2\theta   \\
 &\frac{\rho^2}{C}\frac{(1-D)}{\sin^2\theta}=\frac{1}{\sin^2\theta}-\frac{a^2}{C}
 \end{align}
we obtain after a long but simple computation: 
\begin{align}
&C\left( \frac{\partial S^r}{\partial r}\right)^2-kr^2-\frac{E^2(r^2+a^2)^2}{C}-\frac{L^2 a^2}{C}+2ELa \frac{\rho^2 D}{C}+L^2+a^2E^2=\\
&=-\left( \frac{\partial S^\theta}{\partial \theta}\right)^2+k a^2 \cos^2\theta +a^2E^2 \cos^2\theta-\frac{L^2\cos^2\theta}{\sin^2\theta} ~.
\end{align}
Notice that $\rho^2D$ and $C$ are independent from $\theta$. By looking carefully at the previous expression, we notice that the left-hand side  is independent from $\theta$, while the right-hand side is independent from $r$. 
Therefore we can define an effective separation Carter constant:
\begin{equation}
 \Tilde{C}\equiv\left( \frac{\partial S^\theta}{\partial \theta}\right)^2- \cos^2\theta\left[a^2(k^2+E^2)-\frac{L^2}{\sin^2\theta}\right].
\label{Carter}
\end{equation}
It follows also that
\begin{equation}
 \Tilde{C}=-C\left( \frac{\partial S^r}{\partial r}\right)^2+kr^2+\frac{E^2(r^2+a^2)^2}{C}+\frac{L^2 a^2}{C}-2ELa \frac{\rho^2 D}{C}-L^2-a^2E^2 ~
\label{Carter1}
\end{equation}
 is the fourth integration constant for the solution $S$, together with $E,L,k$. Let's write the geodesic equations in terms of the Carter constant. By defining:
\begin{equation}
 \Theta\equiv\left( \frac{\partial S^\theta}{\partial \theta}\right)^2
\end{equation}
it follows from \eqref{Carter} that:
\begin{equation}
\Theta(\theta)=\Tilde{C}+\cos^2\theta\left[a^2(k^2+E^2)-\frac{L^2}{\sin^2\theta}\right] ~.
\end{equation}
By calling instead
\begin{equation}
 R(r)\equiv\left( \frac{\partial S^r}{\partial r}\right)^2 C^2 ~,
\end{equation}
 from \eqref{Carter1} follows that
\begin{equation}
  R(r)=C\left[-\Tilde{C}+k r^2-(L-aE)^2+(E(r^2+a^2)-La)^2  \right] ~. 
\end{equation}
By plugging these results in \eqref{action} we find
\begin{equation}
 S=-\frac{1}{2}k\lambda-Et+L\psi+\int \sqrt{\Theta (\theta)}\dd \theta+\int \frac{\sqrt{R(r)}}{C}\dd r   ~.
\end{equation}
Now, since
\begin{equation}
 p_{\theta}=\pm \frac{\partial S^\theta}{\partial\theta}=\pm \sqrt{\Theta}=g_{\theta \theta}\Dot{\theta}  ~, 
\end{equation}
the equation of motion for $\theta$ reads
\begin{equation}
\Dot{\theta}=\pm \frac{\sqrt{\Theta}}{\rho^2}    ~;
\label{thetaeq}
\end{equation}
similarly, since
\begin{equation}
 p_r=\pm \frac{\partial S^r}{\partial r}=\pm \frac{\sqrt{R}}{C}=g_{rr}\Dot{r}   
\end{equation}
we get
\begin{equation}
 \Dot{r}=\pm \frac{\sqrt{R}}{\rho^2} ~.   \label{req}
\end{equation}
\eqref{thetaeq} and \eqref{req} together with any two of the equations in \eqref{sistema} give the geodesic equations for this spacetime, both for null ($k=0$), time-like ($k=-1$) or space-like ($k=1$) trajectories.

\acknowledgments
I would like to thank Edward Wilson-Ewing and Lorenzo Cipriani for helpful comments on the manuscript.
This work is supported in part by the Natural Sciences and Engineering Research Council of Canada.


\end{document}